\pdfoutput=1
\documentclass[runningheads,a4paper]{llncs}

\usepackage[T1]{fontenc}
\usepackage{lmodern}
\usepackage{amsmath,amssymb}
\usepackage{booktabs}
\usepackage{array}
\usepackage{graphicx}
\usepackage{xcolor}
\usepackage{tikz}
\usetikzlibrary{arrows.meta,decorations.pathreplacing}
\usepackage{microtype}
\usepackage{xurl}
\usepackage[hidelinks]{hyperref}
\hypersetup{pdftitle={NEST: Stale-Robust Tokenized-Stock Measurement for a Crypto-Treasury LETF Stack},pdfauthor={Doohwi Cha, Minsuk Sung, Minjae Lee, Hyeonjun Yeo, Hanhee Kim, Seunghan Son, Donghwa Seo, Juyeong Lee},pdfkeywords={tokenized stocks, asynchronous markets, stale prices, partial R2, leveraged ETFs}}

\definecolor{nestblue}{HTML}{2A78D6}
\definecolor{nestorange}{HTML}{EB6834}
\definecolor{nestgray}{HTML}{52514E}
\definecolor{nestaxis}{HTML}{9C9B95}

\newcommand{\PDI}{\mathrm{PDI}}
\newcommand{\TableFont}{\footnotesize}
\newcommand{\declaration}[1]{\par\medskip\noindent\textbf{#1}\enspace}

\begin{document}

\title{NEST: Stale-Robust Tokenized-Stock Measurement for a
Crypto-Treasury LETF Stack\thanks{Extended version of a paper accepted at MARBLE 2026
(Lecture Notes in Operations Research, Springer Nature); the changes are listed
in the Declarations.}}
\titlerunning{NEST: Stale-Robust Tokenized-Stock Measurement}

\author{Doohwi~Cha\inst{1}\orcidID{0009-0000-2513-6965}
\and Minsuk~Sung\inst{2}\orcidID{0009-0007-2282-6873}
\and Minjae~Lee\inst{3}\orcidID{0000-0002-8937-9605}
\and Hyeonjun~Yeo\inst{4}\orcidID{0009-0005-9368-306X}
\and Hanhee~Kim\inst{3}\orcidID{0009-0008-1280-7043}
\and Seunghan~Son\inst{3}\orcidID{0009-0000-2902-9999}
\and Donghwa~Seo\inst{5}\orcidID{0009-0003-5718-1535}
\and Juyeong~Lee\inst{6}\orcidID{0000-0002-4619-925X}}
\authorrunning{D. Cha et al.}
\institute{Mirae Asset Securities, Seoul, South Korea; \email{doohwicha@gmail.com}
\and Korea University, Seoul, South Korea; \email{minsuksung@korea.ac.kr}
\and Independent Researcher, Seoul, South Korea;
\email{abraxasnz13@gmail.com, crypt0xb0@gmail.com, qzpm5n@g.skku.edu}
\and Seoul National University, Seoul, South Korea; \email{hj.yeo30@gmail.com}
\and DS Investment \& Securities, Seoul, South Korea;
\email{properitas95@kaist.ac.kr}
\and EY Consulting, Seoul, South Korea; \email{juyeong.lee0839@gmail.com}}

\maketitle

\begin{abstract}
Tokenized equities trade while listed markets are closed, so their marks could
improve off-hours measurement along layered exposures that link Bitcoin,
crypto-treasury stocks, and leveraged ETFs. A token return and the next opening
gap, however, share the price move from the previous close to the token mark,
so part of their apparent fit is mechanical. We develop NEST, a stale-robust
protocol that separates this shared-history fit from conditional measurement
value. A shared-history identity and a variance-clock bound show how much fit
timing alone can generate. The partial determination index (PDI), the squared
partial correlation between token and gap given public controls, measures what
remains, and a finite-sample frontier expresses the partial-$F$ decision in PDI
units. Product, provenance, timestamp, and execution audits decide whether an
informative mark is admissible. For MSTRx, a tracker certificate on
MicroStrategy (MSTR), a token-only $R^2$ of 0.916 shrinks to a modest increment once Bitcoin, mark age,
and a pre-09:00 MSTR premarket proxy are controlled: the token explains 15.7\%
of the remaining gap variation (1\% frontier 0.031), and 14.0\% to 17.0\%
under stricter clock variants, including one in which every token mark
precedes 09:00. A trade-level reconstruction on Gate gives 25.2\%;
NVDAX and AAPLX clear their frontiers, whereas TSLAX does not. An admitted mark
updates, rather than replaces, BTC-unit exposure diagnostics for the
BTC--MSTR--LETF chain.
\keywords{Tokenized stocks \and Asynchronous markets \and Stale prices \and
Partial $R^2$ \and Leveraged ETFs}
\end{abstract}

%======================================================================
\section{Introduction}
%======================================================================

Tokenized equities extend observable prices beyond listed trading hours. A
tracker certificate on a U.S. stock can trade on crypto venues overnight, on
weekends, and on exchange holidays, when the listed share does not trade at
all. Whether such off-hours marks carry information is a practical question for
anyone who monitors exposure while the cash market is shut. We study it for a
particularly levered case, the off-hours MSTRx mark in the chain that links
Bitcoin, a crypto-treasury stock, and a leveraged ETF,
\begin{equation}
 C_t\;(\mathrm{BTC})\longrightarrow
 \mathcal{E}_t=m_t(q_tC_t-D_t)\;(\mathrm{MSTR})\longrightarrow
 L_t\;(\beta\text{-LETF}),                                   \label{eq:chain}
\end{equation}
where $C_t$ is the Bitcoin price, $q_t$ the firm's Bitcoin holdings, $D_t$
its debt, $m_t$ the market premium over net Bitcoin value, and $L_t$ a
leveraged ETF that resets daily to $\beta$ times MSTR's return. The balance
sheet and the premium transmit Bitcoin moves to MSTR, and the daily reset
magnifies the MSTR response. The question is whether MSTRx adds information
beyond Bitcoin and MSTR themselves.

\begin{figure}[t]
\centering
\begin{tikzpicture}[x=1cm,y=1cm,font=\scriptsize]
  \def\close{0.45}\def\mark{6.45}\def\bar{7.85}\def\cut{9.85}\def\open{11.45}
  % time axis with a break for the overnight or weekend interval
  \draw[nestaxis,line width=0.5pt] (0.05,0) -- (2.25,0);
  \draw[nestaxis,line width=0.5pt] (2.85,0) -- (11.9,0);
  \draw[nestaxis] (2.20,-0.11) -- (2.34,0.11);
  \draw[nestaxis] (2.76,-0.11) -- (2.90,0.11);
  \node[text=nestgray,align=center] at (2.55,-0.40) {overnight\\or weekend};
  \foreach \x/\t/\u in {\close/16:00/{close, $d{-}1$},\mark/08:59/{token mark},\bar/09:00/{bar end},%
                        \cut/09:25/{cutoff},\open/09:30/{open, $d$}}{
    \draw[nestaxis,line width=0.5pt] (\x,-0.07) -- (\x,0.07);
    \node[anchor=north,text=nestgray,align=center,inner sep=1pt] at (\x,-0.10) {\t\\\u};}
  % braces above the axis
  \draw[decorate,decoration={brace,amplitude=4pt},line width=0.5pt] (\close,0.16) -- (\mark,0.16)
    node[midway,above=5pt] {$Y_d$: move through the token mark (shared by $G_d$ and $T_d$)};
  \draw[decorate,decoration={brace,amplitude=4pt},line width=0.5pt] (\mark,0.16) -- (\open,0.16)
    node[midway,above=5pt] {$Z_d$: post-mark move};
  % mark age
  \draw[{Stealth[length=3pt]}-{Stealth[length=3pt]},nestgray,line width=0.4pt] (\mark,-0.98) -- (\cut,-0.98);
  \node[text=nestgray,anchor=north,inner sep=1pt] at ({(\mark+\cut)/2},-1.02) {mark age $A_d$ (median 26 min)};
  % information cutoff
  \draw[nestgray,densely dashed,line width=0.4pt] (\cut,-1.38) -- (\cut,-3.75);
  % return intervals, each labelled above its bar
  \draw[black,line width=1.1pt] (\close,-1.85) -- (\open,-1.85);
  \node[anchor=south west,inner sep=1pt] at (\close,-1.82) {$G_d=Y_d+Z_d$: MSTR previous close to open (target)};
  \draw[nestblue,line width=1.1pt] (\close,-2.45) -- (\mark,-2.45);
  \node[anchor=south west,inner sep=1pt,text=nestblue!75!black] at (\close,-2.42) {$T_d=Y_d+\varepsilon_d$: MSTRx return to its mark};
  \draw[nestorange,line width=1.1pt] (\close,-3.05) -- (\bar,-3.05);
  \node[anchor=south west,inner sep=1pt,text=nestorange!75!black] at (\close,-3.02) {$E_d=Y_d+v_d$: MSTR premarket proxy (08:00-ET bar close)};
  \draw[nestgray,line width=1.1pt] (\close,-3.65) -- (\cut,-3.65);
  \node[anchor=south west,inner sep=1pt,text=nestgray] at (\close,-3.62) {$B_d$: Bitcoin return to the cutoff};
  \foreach \y/\x/\c in {-1.85/\open/black,-2.45/\mark/nestblue,-3.05/\bar/nestorange,-3.65/\cut/nestgray}{
    \fill[\c] (\x,\y) circle (1.4pt);}
\end{tikzpicture}
\caption{Information clock for one closed-market window (schematic, not to
scale). The opening gap $G_d$ and the token return $T_d$ share the move $Y_d$
from the previous close to the token mark, which raises raw fit mechanically
(Proposition~\ref{prop:shared}). NEST asks whether $T_d$ still explains $G_d$
after conditioning on information observable by the 09:25 cutoff: the Bitcoin
return $B_d$, the mark age $A_d$, and MSTR's premarket proxy $E_d$. In the
218-window premarket sample the composite mark falls between 08:56 and 09:15 ET,
and 214 of the 218 marks precede 09:00.}
\label{fig:clock}
\end{figure}
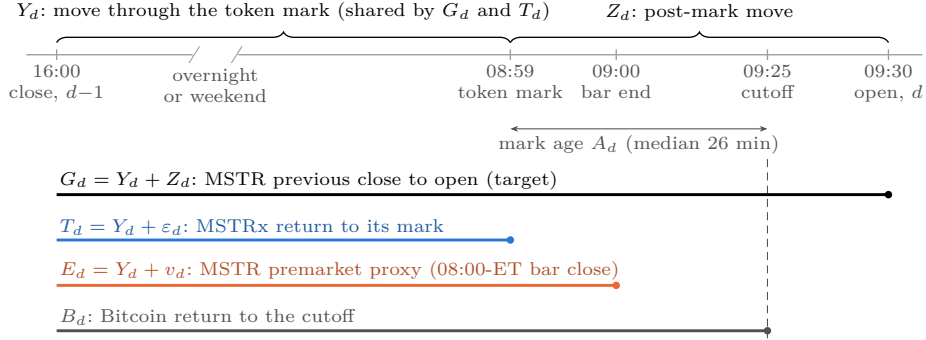

High raw fit does not answer that question. Both the token return and the next
opening gap start at (or, for the token, near) the previous MSTR close, so they
share the previous-close-to-mark price move (Fig.~\ref{fig:clock}). That common prefix
raises fit mechanically, even when the token says nothing about the move after
its mark. Three estimands are therefore easy to conflate: whether the token
reflects information by its mark (\emph{interval incorporation}), whether it
explains the gap beyond public information observable at the cutoff
(\emph{conditional informativeness}), and whether it predicts later shocks
(\emph{post-mark forecasting}). NEST targets the second, after characterizing
the first.

NEST (Nested Exposure Slope and Thresholds) is a stale-robust admission
protocol. A shared-history identity and a variance-clock bound quantify the fit
that timing alone can create. The partial determination index (PDI) and its
finite-sample frontier then test what the mark adds beyond public inputs
observable by the cutoff---Bitcoin, the age of the mark, and MSTR's own premarket price. Product
rights, price provenance, timestamps, and execution costs form the
admissibility layer that decides whether a statistical increment may be used.
An admitted mark then updates BTC-unit exposure diagnostics, including the
Bitcoin move that would extinguish a daily-reset LETF in one interval.

The empirical answer is a modest, complementary increment. On 222
strict-cutoff windows from July 2025 to June 2026, the token-only regression
has $R^2=0.916$. MSTR's pre-09:00 premarket proxy absorbs most of that fit, yet
MSTRx still explains 15.7\% of the gap variation left by Bitcoin, mark age, and
the proxy ($\PDI=0.157$; HAC $t=4.61$; 1\% frontier 0.031). The increment is
14.0\% when every token mark is required to precede the proxy's 09:00 bar end,
25.2\% for a trade-level reconstruction on Gate, and statistically the same on
weekdays and weekends. Across other xStocks it is product specific: NVDAX and
AAPLX clear their frontiers and TSLAX does not.

The paper makes three contributions.
\begin{enumerate}
\item \emph{Conceptual.} It shows that shared timing, not information, can
generate most of the raw fit between an off-hours mark and the next opening gap,
and it bounds that fit by the share of closed-window variance realized by the
mark (Propositions~\ref{prop:shared}--\ref{prop:clock}).
\item \emph{Measurement.} It recasts the conditional question as a partial
$R^2$ with an exact finite-sample frontier and power function, gives the
interpretation of a positive PDI against a noisy public benchmark, and implements a same-clock
design whose timestamps cannot see the open
(Lemmas~\ref{lem:pdi}--\ref{lem:noisy}, Theorem~\ref{thm:frontier},
Corollary~\ref{cor:mde},
Sects.~\ref{sec:data}--\ref{sec:results}).
\item \emph{Operational.} It audits product rights, price provenance, a named
venue, and execution cost, and it translates an admitted mark into BTC-unit
exposure and one-reset wipeout thresholds for the LETF layer
(Sect.~\ref{sec:stack}, Table~\ref{tab:stack}).
\end{enumerate}

Section~\ref{sec:lit} relates NEST to the price-discovery and tokenized-asset
literatures. Section~\ref{sec:model} develops the measurement model and
Section~\ref{sec:stack} the stack and admissibility diagnostics.
Section~\ref{sec:data} describes the data and the same-clock design,
Section~\ref{sec:results} reports the results, and
Sections~\ref{sec:discussion}--\ref{sec:conclusion} discuss implications,
limitations, and conclusions.

%======================================================================
\section{Related Literature}
\label{sec:lit}
%======================================================================

\paragraph{Price discovery across markets.} Information shares and component
shares allocate innovations in a common efficient price across cointegrated
markets \cite{hasbrouck1995one,gonzalo1995estimation}. Putni{\c{n}}{\v{s}}
\cite{putnins2013what} shows that these metrics mix information leadership with
differences in noise, so a market can rank highly by being less noisy rather
than by moving first. NEST does not allocate a common price. It asks whether one
intermittent mark adds explanatory power for a later target, conditional on
measured public information, and its noisy-control lemma makes the noise
channel explicit.

\paragraph{Asynchronous and sequential trading.} Nonsynchronous trading biases
betas and return moments \cite{scholes1977estimating,lo1990econometric}.
When trading hours do not overlap, state-space models recover around-the-clock
price discovery from noncontemporaneous quotes \cite{menkveld2007modelling};
sequential-market measures \cite{wang2011information} and information shares
for partially overlapping hours \cite{dimpfl2023price} adapt the
price-discovery toolkit to markets that open one after another; and asynchronous
updates can overstate the discovery attributed to one market
\cite{han2025asynchronous}. The closest antecedent in spirit is the evidence
that after-hours and pre-open trading in U.S. stocks is informative despite low
volume \cite{barclay2003price}. Interval-$R^2$ measures how much of a return is
incorporated within a subperiod \cite{vanbommel2011measuring}. NEST differs in
the object it evaluates: an aggregator-produced reference mark whose return
shares a prefix with its target, which makes raw fit uninformative until the
shared history is diagnosed.

\paragraph{Tokenized stocks.} Cong et al.\ \cite{cong2025tokenized} document
that weekend token moves anticipate subsequent stock openings, and Dombrowski
and Arshadi \cite{dombrowski2026tokenized} report weak tracking and arbitrage
limits when listed markets are closed. A contemporaneous daily-frequency study
of matched xStock--underlying pairs attributes most long-run price discovery to
the listed equity \cite{lin2026one}. These results leave open the conditional
contribution of an off-hours mark once its shared timing and the listed stock's
own premarket price are taken into account; NEST supplies a test of exactly
that contribution together with the audits needed to use it.

\paragraph{Leveraged ETFs.} Daily-reset funds compound path-dependently, and
their returns over longer horizons decay with realized variance
\cite{cheng2009dynamics,avellaneda2010path}. In the stack (\ref{eq:chain}) the
LETF's underlier is itself a balance-sheet-levered claim on Bitcoin, which is
why an off-hours measurement of MSTR's Bitcoin exposure matters for the LETF
layer.

%======================================================================
\section{Measurement Model}
\label{sec:model}
%======================================================================

Table~\ref{tab:notation} aligns each variable with the time at which it is
observed. Returns are logarithmic unless stated otherwise.

\begin{table}[t]
\centering
\TableFont
\caption{Core variables and their measurement times.}
\label{tab:notation}
\begin{tabular*}{\textwidth}{@{\extracolsep{\fill}}lp{9.6cm}@{}}
\toprule
Symbol & Meaning and timing\\
\midrule
$G_d$ & MSTR opening gap: previous official close to the 09:30 ET open of day $d$\\
$T_d$ & token return from the mark nearest the previous 16:00 ET close to the last mark at or before 09:25 ET\\
$B_d$, $A_d$ & Bitcoin return over the same window; age of the token end mark at the 09:25 cutoff\\
$E_d$ & MSTR premarket proxy: close of the 08:00-ET hourly bar relative to the previous close\\
$Y_d$, $Z_d$, $\varepsilon_d$ & move through the token mark; later move to the open; mark-specific noise\\
$S$ & prespecified non-token controls: $(1,B_d,A_d)$, or $(1,B_d,A_d,E_d)$ in the premarket test\\
$\PDI_{T\mid S}$ & partial determination index: squared partial correlation of $G$ and $T$ given $S$\\
$\kappa$, $\beta$ & effective MSTR slope on Bitcoin; LETF daily target leverage\\
\bottomrule
\end{tabular*}
\end{table}

\subsection{Mechanical Fit from Shared History}

Write the opening gap as $G_d=Y_d+Z_d$ and the token return as
$T_d=Y_d+\varepsilon_d$, where $Y_d$ is the move through the token mark, $Z_d$
the subsequent move to the open, and $\varepsilon_d$ mark-specific measurement
noise from stale constituents, aggregation, or microstructure.

\begin{proposition}[Shared-history identity]\label{prop:shared}
Suppose $Y_d,Z_d,\varepsilon_d$ are mean zero and pairwise uncorrelated, with
variances $\sigma_Y^2,\sigma_Z^2,\sigma_\varepsilon^2$, and that
$\operatorname{Var}(G_d)$ and $\operatorname{Var}(T_d)$ are positive. The
population regression of $G_d$ on $T_d$ has slope, fit, and post-mark
covariance
\begin{equation}
 b_T=\frac{\sigma_Y^2}{\sigma_Y^2+\sigma_\varepsilon^2},\qquad
 R_T^2=\frac{\sigma_Y^4}
 {(\sigma_Y^2+\sigma_Z^2)(\sigma_Y^2+\sigma_\varepsilon^2)},\qquad
 \operatorname{Cov}(Z_d,T_d)=0.                         \label{eq:stale}
\end{equation}
\end{proposition}
\begin{proof}
Pairwise orthogonality gives $\operatorname{Cov}(G,T)=\sigma_Y^2$,
$\operatorname{Var}(G)=\sigma_Y^2+\sigma_Z^2$, and
$\operatorname{Var}(T)=\sigma_Y^2+\sigma_\varepsilon^2$. Substituting into
the population slope $b_T=\operatorname{Cov}(G,T)/\operatorname{Var}(T)$ and the
squared correlation gives the first two expressions, and
$\operatorname{Cov}(Z,Y+\varepsilon)=0$ gives the third. \qed
\end{proof}

Raw fit can therefore be high even though the token has zero covariance with
everything that happens after its mark: fit is governed by how large the shared
move is relative to the post-mark move and to mark noise. Because volatility
does not accrue uniformly in calendar time---it is concentrated around news and
around the U.S.\ premarket---the relevant timing screen uses a variance clock.

\begin{proposition}[Variance-clock bound]\label{prop:clock}
Take the closed window as $[0,1]$. Let $M_t$ be a square-integrable process with
orthogonal increments and variance clock $V(t)=\operatorname{Var}(M_t-M_0)$.
For a mark time $\theta\in[0,1]$, set $G=M_1-M_0$ and
$T=M_\theta-M_0+\varepsilon$, where $\varepsilon$ is mean zero, has variance
$\sigma_\varepsilon^2$, and is uncorrelated with the increments of $M$. If
$V(1)>0$ and $V(\theta)+\sigma_\varepsilon^2>0$, then
\begin{equation}
 R^2(G\mid T)
 =\frac{V(\theta)^2}{V(1)\{V(\theta)+\sigma_\varepsilon^2\}}
 \leq \frac{V(\theta)}{V(1)},\qquad
 \operatorname{Cov}(M_1-M_\theta,T)=0.                  \label{eq:clock}
\end{equation}
\end{proposition}
\begin{proof}
Set $Y=M_\theta-M_0$ and $Z=M_1-M_\theta$. Orthogonal increments give
$\operatorname{Cov}(Y,Z)=0$, $\operatorname{Var}(Y)=V(\theta)$, and
$\operatorname{Var}(G)=V(1)$, so Proposition~\ref{prop:shared} gives the
equality. The bound follows from $V(\theta)/\{V(\theta)+\sigma_\varepsilon^2\}\le 1$. \qed
\end{proof}

The bound $V(\theta)/V(1)$ is the share of closed-window variance realized by
the time of the mark. A mark that arrives late in the variance clock can fit the
gap almost perfectly without any information about the remainder, so the bound
is a timing screen, not evidence of measurement value. Square-integrable
martingales, such as efficient log prices, satisfy the orthogonal-increment
condition.

\subsection{Conditional Measurement: the Partial Determination Index}

Let $S$ be an $n\times p$ full-rank matrix containing an intercept and the
prespecified non-token controls, let $M_S=I-S(S'S)^{-1}S'$ be the residual
maker, and define $R_0^2=R^2(G\mid S)$ and $R_1^2=R^2(G\mid S,T)$.

\begin{lemma}[PDI decomposition]\label{lem:pdi}
If $M_SG\ne0$ and $M_ST\ne0$, then
\begin{equation}
 \PDI_{T\mid S}
 \equiv\operatorname{Corr}^2(M_SG,M_ST)
 =\frac{R_1^2-R_0^2}{1-R_0^2},                                      \label{eq:pdi}
\end{equation}
so that $\Delta R^2\equiv R_1^2-R_0^2=(1-R_0^2)\,\PDI_{T\mid S}$.
\end{lemma}
\begin{proof}
Let $\widetilde G=M_SG$ and $\widetilde T=M_ST$. The restricted residual sum of
squares is $RSS_0=\widetilde G'\widetilde G$. By the Frisch--Waugh--Lovell
theorem \cite{frisch1933partial,lovell1963seasonal}, adding $T$ lowers it by
$(\widetilde G'\widetilde T)^2/(\widetilde T'\widetilde T)$, so
$(RSS_0-RSS_1)/RSS_0=\operatorname{Corr}^2(\widetilde G,\widetilde T)$. With an
intercept in $S$, $RSS_0=(1-R_0^2)\,TSS$ and $RSS_1=(1-R_1^2)\,TSS$, which gives
(\ref{eq:pdi}). \qed
\end{proof}

The PDI is the classical coefficient of partial determination. It measures the
share of gap variation left unexplained by $S$ that the token captures, and
$\Delta R^2$ converts it back into an absolute fit gain. Both are conditional
on the measured controls; neither identifies which market originated the
information. The next result shows how a noisy public benchmark shapes the
interpretation.

\begin{lemma}[Noisy-control interpretation]\label{lem:noisy}
Let $G=Y+Z$, $T=Y+u$, and $E=Y+v$ with centered, pairwise-orthogonal
components, and write $(a,b,c,d)=(\operatorname{Var}Y,\operatorname{Var}Z,
\operatorname{Var}u,\operatorname{Var}v)$ with $a,b,c>0$ and $d\geq0$. With
$h=ad/(a+d)$,
\begin{equation}
 \PDI_{T\mid E}=\frac{h^2}{(b+h)(c+h)}.           \label{eq:noisy}
\end{equation}
It is zero at $d=0$, increases with the control noise $d$, decreases in $b$ and
$c$, and converges to $a^2/\{(a+b)(a+c)\}$ as $d\to\infty$.
\end{lemma}
\begin{proof}
Projecting on $E$ removes $\{a/(a+d)\}E$ from both $G$ and $T$. The residual
covariance is $a-a^2/(a+d)=h$, and the residual variances are
$a+b-a^2/(a+d)=b+h$ and $c+h$, which gives (\ref{eq:noisy}). Since
$\partial\log\PDI/\partial h=2/h-1/(b+h)-1/(c+h)>0$ and $h$ increases in $d$
from $0$ to $a$, the comparative statics follow. \qed
\end{proof}

A positive PDI against a noisy benchmark therefore means improvement over that
benchmark, not exclusive information origin: if the proxy observed the shared
move without noise ($d=0$), a token with no post-mark information would have
$\PDI=0$. Sampling uncertainty is expressed on the same scale.

\begin{theorem}[Finite-sample PDI frontier and power]\label{thm:frontier}
Assume that $n>p+1$, that $[S\;T]$ has rank $p+1$, and that
$G=S\gamma+T\delta+\eta$ with $\eta\mid S,T\sim N(0,\sigma^2I_n)$. Let
$\widehat{\PDI}_{T\mid S}$ denote the sample version of Lemma~\ref{lem:pdi},
let $\nu=n-p-1$, and define the signal share
$\theta=\delta^2T'M_ST/(\delta^2T'M_ST+n\sigma^2)\in[0,1)$. Then
\begin{equation}
 F_{\PDI}
 =\frac{\nu\,\widehat{\PDI}_{T\mid S}}
 {1-\widehat{\PDI}_{T\mid S}}
 \sim F'_{1,\nu}\!\Big(\frac{n\theta}{1-\theta}\Big),\qquad
 c_\alpha(n,p)=\frac{F_{1,\nu;1-\alpha}}
 {\nu+F_{1,\nu;1-\alpha}},                              \label{eq:frontier}
\end{equation}
where $F'_{1,\nu}(\lambda)$ is the noncentral $F$ distribution with
noncentrality $\lambda$. Under $H_0:\delta=0$, equivalently $\theta=0$, it is
the central $F_{1,\nu}$, and the level-$\alpha$ partial-$F$ test rejects exactly
when $\widehat{\PDI}_{T\mid S}>c_\alpha(n,p)$.
\end{theorem}
\begin{proof}
By Lemma~\ref{lem:pdi}, $\widehat{\PDI}=(RSS_0-RSS_1)/RSS_0$, hence
$\nu\widehat{\PDI}/(1-\widehat{\PDI})=(RSS_0-RSS_1)/(RSS_1/\nu)$. Conditional
on $(S,T)$, $\widetilde G=M_SG=\widetilde T\delta+M_S\eta$ with
$\widetilde T=M_ST$, so $RSS_0-RSS_1=(\widetilde T'\widetilde G)^2/(\widetilde
T'\widetilde T)$ is $\sigma^2$ times a noncentral $\chi^2_1$ variable with
noncentrality $\delta^2T'M_ST/\sigma^2=n\theta/(1-\theta)$, while
$RSS_1/\sigma^2\sim\chi^2_\nu$ is independent of it because the two quadratic
forms project $\eta$ onto orthogonal subspaces. Their scaled ratio is therefore
$F'_{1,\nu}(n\theta/(1-\theta))$. Since $x\mapsto\nu x/(1-x)$ is increasing on
$[0,1)$, $F_{\PDI}>F_{1,\nu;1-\alpha}$ is equivalent to
$\widehat{\PDI}>c_\alpha(n,p)$. \qed
\end{proof}

\begin{corollary}[Minimum detectable PDI]\label{cor:mde}
Under the assumptions of Theorem~\ref{thm:frontier}, the power of the
level-$\alpha$ test, $\Pr\{F'_{1,\nu}(n\theta/(1-\theta))>F_{1,\nu;1-\alpha}\}$,
is continuous and strictly increasing in $\theta$, rising from $\alpha$ toward
one. For every target power $\pi\in(\alpha,1)$ there is therefore a unique
minimum detectable signal share $\theta_\pi(n,p,\alpha)$, and for large $\nu$,
$\theta_\pi\approx K^2/(n+K^2)$ with $K=z_{1-\alpha/2}+z_\pi$.
\end{corollary}
\begin{proof}
The noncentral $F$ distribution is stochastically increasing in its
noncentrality, which is continuous and increasing in $\theta$ and diverges as
$\theta\to1$; this gives existence and uniqueness. As $\nu\to\infty$, the
signed square root of the numerator is $N(\sqrt\lambda,1)$, so power is
approximately $\Phi(\sqrt\lambda-z_{1-\alpha/2})$; equating it to $\pi$ gives
$\lambda=K^2$ and $\theta_\pi=K^2/(n+K^2)$. \qed
\end{proof}

Theorem~\ref{thm:frontier} restates the classical nested-model test in PDI
units and adds its power function; the signal share $\theta$ is the population
counterpart of the PDI. The frontier lets a reader compare an estimated share
directly with the share required for significance, and Corollary~\ref{cor:mde}
turns it into a design tool. At the 1\% level and 80\% power, 218 windows detect
any PDI above 0.052, whereas 48 windows detect only PDIs above 0.21, so a small
subsample that fails its frontier is uninformative unless its PDI is compared
with $\theta_{80}$. Because daily return errors are heteroskedastic and possibly
autocorrelated, every frontier decision is also paired with a Newey--West HAC
$t$ statistic with five lags on the token coefficient \cite{newey1987simple}.

\subsection{The Admission Protocol}

NEST admits an off-hours mark as a measurement input in four steps.
(i)~\emph{Clock}: every input must be observable by the cutoff, and the
shared-history and variance-clock diagnostics are reported next to raw fit.
(ii)~\emph{Conditioning}: the prespecified controls $S$ include the strongest
cutoff-observable public benchmark, here MSTR's own premarket price.
(iii)~\emph{Decision}: the mark clears when the estimated PDI exceeds the
frontier $c_{.01}(n,p)$ and the HAC check agrees. (iv)~\emph{Admissibility}: product
rights, price provenance, a named-venue replication, and execution costs
determine how far a cleared mark can be used
(Sect.~\ref{sec:stack}). The protocol does not ask the token to be the
dominant venue; it asks whether the mark is a reliable, conditional input.

%======================================================================
\section{Stack Exposure and Admissibility}
\label{sec:stack}
%======================================================================

An admitted mark is useful only if it can be mapped into the exposures that the
stack (\ref{eq:chain}) creates. Let $g_k$ be MSTR's simple return in reset
interval $k$, $\beta$ the LETF's target leverage, and $f$ its fee rate per
interval.

\begin{lemma}[Daily-reset identity]\label{lem:reset}
On paths with $1+\beta g_k>0$ for every $k$,
\begin{equation}
 \begin{aligned}
 \frac{L_N}{L_0}&=\prod_{k=1}^N(1+\beta g_k)\,e^{-fN},\\
 \log\frac{L_N}{L_0}&=\beta\log\frac{\mathcal E_N}{\mathcal E_0}
 -\frac{\beta(\beta-1)}2\sum_k g_k^2-fN+O\Big(\sum_k|g_k|^3\Big).
 \end{aligned}                                                     \label{eq:reset}
\end{equation}
\end{lemma}
\begin{proof}
The product is the definition of a daily reset with a continuously accrued fee.
Expanding $\log(1+\beta g)=\beta g-\beta^2g^2/2+O(|g|^3)$ and
$\beta\log(1+g)=\beta g-\beta g^2/2+O(|g|^3)$, and summing over $k$, gives the
second expression \cite{cheng2009dynamics,avellaneda2010path}. \qed
\end{proof}

For leveraged or inverse funds ($\beta>1$ or $\beta<0$) the quadratic term is a
variance drag. The equity layer adds balance-sheet leverage. With
$\mathcal{E}=m(qC-D)$ and a locally linear premium response
$d\log m=\rho_m\,d\log C+e$, the effective slope of MSTR on Bitcoin is
\begin{equation}
 \kappa=\lambda_{\mathcal{E}}+\rho_m,\qquad
 \lambda_{\mathcal{E}}=\frac{qC}{qC-D},\qquad
 \frac{\partial\lambda_{\mathcal E}}{\partial C}=-\frac{qD}{(qC-D)^2}<0, \label{eq:kappa}
\end{equation}
for $q,D>0$ and $qC>D$. A Bitcoin drawdown therefore raises book leverage mechanically, and the premium can
reinforce or offset that channel.

\begin{corollary}[One-reset wipeout threshold]\label{cor:wipe}
If the premium is constant over a reset interval and $q,D$ are fixed, the
Bitcoin move that sets a $\beta$-LETF to zero in that interval is
$(\Delta C/C)_{\rm wipe}=-1/(\beta\lambda_{\mathcal E})$. Replacing
$\lambda_{\mathcal E}$ by $\kappa$ gives the first-order comparison
$-1/(\beta\kappa)$ when the premium moves with Bitcoin.
\end{corollary}
\begin{proof}
With constant $m$, $\Delta\mathcal E/\mathcal E=q\Delta C/(qC-D)
=\lambda_{\mathcal E}\,\Delta C/C$ exactly, and the LETF value reaches zero when
$1+\beta g=0$. With a moving premium, $g\approx\kappa\,\Delta C/C$ to first
order. \qed
\end{proof}

The threshold is a link-specific exposure diagnostic: it depends on which
estimate of $\kappa$---book leverage, the listed-session beta, or the slope
implied by an admitted off-hours mark---is used to update it
(Sect.~\ref{sec:res-stack}).

\paragraph{Executable band.} Statistical informativeness does not make a token
price enforceable. Let $P_t$ be the per-unit token price, $V_t^{\rm ref}$ the
per-unit reference value of the underlying, $\tau$ the settlement delay, $r$
the riskless rate, and $k^-(\tau,x),k^+(\tau,x)$ the present-value per-unit
all-in costs of the buy--redeem and borrow--sell--create strategies at size
$x$, with $k^+$ including borrowing and funding.

\begin{proposition}[Executable band conditional on access]\label{prop:band}
Suppose an eligible trader can redeem size $x$, and can borrow and sell $x$
tokens at $t$ while committing to create the same size for delivery at
$t+\tau$, with both requests locking $V_t^{\rm ref}$ at $t$. Absence of
arbitrage then implies
\begin{equation}
 e^{-r\tau}V_t^{\rm ref}-k^-(\tau,x)\leq P_t\leq
 e^{-r\tau}V_t^{\rm ref}+k^+(\tau,x).                    \label{eq:band}
\end{equation}
\end{proposition}
\begin{proof}
Below the lower bound, buying and redeeming earns a positive present value after
redemption costs; above the upper bound, borrowing and selling now and covering
with the locked creation at $t+\tau$ does so after creation, borrowing, and
funding costs. Either trade contradicts absence of arbitrage. \qed
\end{proof}

Without the stated locking, eligibility, and inventory, (\ref{eq:band}) is only
a friction benchmark. Creation and redemption of MSTRx are restricted by KYC,
whitelisting, fees, and minimum sizes (Sect.~\ref{sec:data}), so the band is
not enforceable for most holders. NEST therefore reports these conditions and a
named-venue execution cost instead of assuming arbitrage.

%======================================================================
\section{Data and Same-Clock Design}
\label{sec:data}
%======================================================================

\paragraph{What the mark represents.} MSTRx is a variable-denomination tracker
certificate issued by Backed Assets (JE) Ltd.; it is not MSTR common stock and
carries no voting rights \cite{backedMSTRx,xstocksFAQ}. The issuer reports
segregated MSTR collateral. On EVM chains the token rebases and on Solana it
uses the Token-2022 Scaled UI extension, so corporate actions are passed through
by scaling rather than by price formation \cite{xstocksDevelopers}. Primary
creation and redemption are subject to KYC, whitelisting, fees, and minimum
sizes. An informative secondary mark is therefore a measurement of a
collateralized tracker outside listed hours, not of unrestricted ownership or of
an enforceable primary-market arbitrage.

\paragraph{Price sources.} The main signal is DefiLlama's archive of
CoinGecko's multi-venue MSTRx reference price \cite{defillamaPrices}. CoinGecko
normalizes venue prices, removes outliers, and applies volume weights, but does
not publish the historical constituents and weights behind each mark
\cite{coingeckoMethodology}. Every observed MSTRx confidence value equals 0.99,
so the confidence field screens availability, not liquidity. Bitcoin returns
use DefiLlama's archived CoinGecko Bitcoin price at the same targets. The target
joins Financial Modeling Prep's previous official MSTR close to the next 09:30
ET official open. To obtain venue evidence that does not depend on the
composite's undisclosed composition, we reconstruct 694,246 MSTRX--USDT trades
from Gate's twelve official monthly trade files for July 2025--June 2026 and
take the last trade at or before each target, at most 15 minutes old; 24
order-book snapshots at the window endpoints provide a cost check
\cite{gateHistoricalQuotes}.

\paragraph{A clock that cannot see the open.} Candidate windows join adjacent
U.S. trading sessions from July 2, 2025 through June 5, 2026. Token and Bitcoin
marks are requested at the previous 16:00 ET close and at a 09:25 ET cutoff,
five minutes before the open. An earlier nearest-mark rule could select prices
observed after the cutoff: of the 232 windows that pass the availability
screens, 10 used a token end mark after 09:25, and 8 of these a mark at or after
the 09:30 open. The primary design
instead requires the token end mark to be available by the cutoff, and a
stricter sensitivity requires both token marks to precede their targets.
Table~\ref{tab:sample} keeps the economically different exclusions visible. The
composite updates roughly hourly, so the end mark is typically the 08:59 print:
its age at the cutoff has median 26.0 minutes and standard deviation 1.5
minutes.

\begin{table}[t]
\centering
\TableFont
\caption{Sample construction for MSTR/MSTRx. Availability screens precede the
backward cutoff repair; the 222-window row is the primary sample.}
\label{tab:sample}
% Generated by scripts/build_arxiv_assets.py from archived results; do not edit.
\begin{tabular*}{\textwidth}{@{\extracolsep{\fill}}lrrl@{}}
\toprule
Screen & Retained & Removed & Role\\
\midrule
Adjacent U.S. session windows & 234 & 0 & Universe\\
All four price endpoints observed & 233 & 1 & Missing token end mark\\
Confidence at least 0.8 & 233 & 0 & Availability screen\\
Endpoint lags at most 60 min & 232 & 1 & Start mark 421 min old\\
\addlinespace
Token end mark by 09:25 ET & 222 & 10 & Primary sample\\
Both token marks by their targets & 219 & 3 & Stricter clock\\
\bottomrule
\end{tabular*}

\end{table}

The two windows removed by the availability screens are calmer than the
retained ones (mean absolute gap 1.50\% versus 1.87\%; mean absolute Bitcoin move
1.08\% versus 1.53\%), and the Spearman correlations of mark lag with the
absolute gap and the absolute Bitcoin move are insignificant ($p=0.655$ and
$0.642$). With two exclusions this diagnostic has little power, so we report it
rather than rely on it. The three windows that start after a 13:00 ET early
close \cite{nyseHours} and the 41 strictly pre-target Friday-to-Monday windows
enter as sensitivities.

\paragraph{The public benchmark the token must beat.} MSTR's own premarket price
is the economically relevant alternative. Historical Financial Modeling Prep
data begin at the regular open, so we archive Yahoo Finance's hourly
extended-hours chart \cite{yahooFinanceChart} and define
$E_d=\log(P^{08}_d/P_{d-1})$, where $P^{08}_d$ is the close of the bar labelled
08:00 ET. Wherever the next bar is observed (217 of the 218 matched windows) it
starts at 09:00, so $P^{08}_d$ is the last value before 09:00. The proxy is coarse: the provider does not document
whether bars are built from trades or quotes, every matched bar reports zero
volume, and four of the 222 strict windows lack an 08:00 bar (their mean absolute
gap is 2.63\% against 1.83\%; Mann--Whitney $p=0.99$). The 09:00 bar close is
excluded because it is mechanically the open (its correlation with the official
opening price exceeds 0.9999). This leaves 218 common windows, on which the token and
the proxy are added in both orders after Bitcoin and mark age. By
Lemma~\ref{lem:noisy}, a surviving token increment is improvement over this
observed proxy and leaves the origin of information open.

\paragraph{Clock audit.} The composite end marks in the 218-window sample fall
between 08:56 and 09:15 ET. In 214 windows they precede 09:00 by 0.85 to 3.15
minutes (median 0.97), so the token mark and the nominal end of the proxy bar
essentially coincide; in the remaining four the mark is 0.2 to 15.8 minutes later than 09:00.
Section~\ref{sec:res-clock} reports the test without those four windows. The
Bitcoin control is measured to the 09:25 cutoff, about 26 minutes after the
token mark; this asymmetry works against the token, because Bitcoin information
arriving after the mark is credited to the controls. Bitcoin end marks are at
most 41 seconds after 09:25 (112 windows) and always precede the open.

\paragraph{Venue and portability designs.} The Gate premarket comparison fixes
the Gate end price at the last trade by 08:59 ET, aligned with the end of the
proxy bar. A same-clock comparison evaluates Gate at the composite's own
timestamps and conditions each source on the other. The cross-name exercise
repeats the premarket-conditioned test for NVDAX, TSLAX, and AAPLX with the same
strictly backward rule and each underlying's own 08:00-ET bar, giving 204, 197,
and 224 windows. Expanding-window and frozen July 2026 evaluations test whether
the broader Bitcoin-and-age increment persists out of sample.

%======================================================================
\section{Results}
\label{sec:results}
%======================================================================

\subsection{From Raw Fit to Conditional Value}

The raw relation shows why the timing diagnostics are needed. On the 222-window
strict sample, the token-only regression has $R^2=0.916$. Relative to Bitcoin
and mark age, MSTRx adds $\Delta R^2=0.133$ and has $\PDI=0.620$ (HAC
$t=10.92$), far above its 1\% frontier of 0.030 (Table~\ref{tab:main}). This
comparison, however, omits the listed stock's own premarket price.

\begin{figure}[t]
\centering
\includegraphics[width=\textwidth]{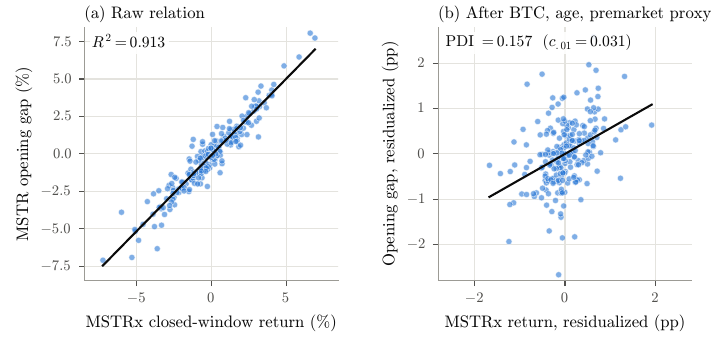}
\caption{Raw and conditional fit on the 218-window premarket sample. (a) The
MSTR opening gap against the MSTRx closed-window return; most of this fit is the
shared previous-close-to-mark move. (b) The same two series after residualizing
both on an intercept, the Bitcoin return, mark age, and the premarket proxy
(Frisch--Waugh--Lovell); the squared correlation of the residuals is the PDI and
the line has the full-model token slope 0.57.}
\label{fig:fit}
\end{figure}

The premarket benchmark is the harder test. On the 218 common windows, Bitcoin
and age explain 77.9\% of the gap variation and adding the premarket proxy raises
this to 90.8\%. MSTRx then adds $\Delta R^2=0.014$, so the full model explains
92.3\%. The increment is small in absolute terms but clears the frontier
decisively: the token explains 15.7\% of the variation left unexplained
($\PDI=0.157$, $F=39.6$, HAC $t=4.61$, $c_{.01}=0.031$).
Figure~\ref{fig:fit} shows the contrast between the raw relation and the
residualized one. Token and proxy are highly correlated (0.970), and the reverse
increment of the proxy given the token is $\PDI=0.082$: the two prices overlap
substantially without either spanning the other. MSTRx is a complementary
off-hours measurement, not a dominant price-discovery venue.

\begin{table}[t]
\centering
\TableFont
\caption{Conditional incremental fit. $R^2_0$ is the fit of the conditioning
set $S$, $\Delta R^2$ the absolute gain from adding the token, and PDI the share
of the remaining variation that the token explains; ``Clears'' means
$\PDI>c_{.01}(n,p)$. The Gate premarket row ends the Gate return at 08:59 ET.
HAC $t$ uses five Newey--West lags.}
\label{tab:main}
% Generated by scripts/build_arxiv_assets.py from archived results; do not edit.
\setlength{\tabcolsep}{4pt}
\begin{tabular*}{\textwidth}{@{\extracolsep{\fill}}lrrrrrrc@{}}
\toprule
Signal; conditioning set $S$ & $N$ & $R^2_0$ & $\Delta R^2$ & PDI & $c_{.01}$ & HAC $t$ & Clears\\
\midrule
\multicolumn{8}{@{}l}{\textit{MSTRx composite}}\\
\quad BTC, age & 222 & 0.785 & 0.133 & 0.620 & 0.030 & 10.92 & yes\\
\quad BTC, age, premarket & 218 & 0.908 & 0.014 & 0.157 & 0.031 & 4.61 & yes\\
\addlinespace
\multicolumn{8}{@{}l}{\textit{Gate trade, MSTRX--USDT}}\\
\quad BTC, age & 225 & 0.778 & 0.176 & 0.792 & 0.030 & 26.78 & yes\\
\quad BTC, age, premarket (08:59) & 216 & 0.908 & 0.023 & 0.252 & 0.031 & 6.85 & yes\\
\addlinespace
\multicolumn{8}{@{}l}{\textit{Other xStocks; $S$: BTC, age, own premarket}}\\
\quad NVDAX & 204 & 0.891 & 0.018 & 0.169 & 0.033 & 4.85 & yes\\
\quad AAPLX & 224 & 0.749 & 0.023 & 0.092 & 0.030 & 5.16 & yes\\
\quad TSLAX & 197 & 0.896 & 0.001 & 0.010 & 0.034 & 0.80 & no\\
\bottomrule
\end{tabular*}

\end{table}

\subsection{Clock Audit of the Premarket Comparison}
\label{sec:res-clock}

Table~\ref{tab:timing} varies the clock of the premarket comparison. Dropping
the four windows whose token mark is later than 09:00 lowers the PDI from 0.157
to 0.140 (HAC $t=4.36$), still well above the frontier of 0.031; in this sample
every token mark precedes the nominal end of the proxy bar by at most about
three minutes.
Restricting to windows whose Bitcoin end mark is at or before 09:25 halves the
sample but leaves the PDI at 0.168. Requiring both token marks to precede their
targets, tightening the matching tolerance, winsorizing, or replacing the proxy
by the open of the 09:00 bar leaves the PDI between 0.150 and 0.170. The
reverse increment of the proxy remains positive throughout, consistent with two
noisy measurements of the same closed-market move.

The increment does not depend on the type of closure. The 170 weekday windows
give $\PDI=0.122$ and the 48 weekend-or-holiday windows 0.112, and in the full
premarket model a token-by-weekend interaction is far from significant (HAC
$t=-0.14$, $p=0.89$; token loadings 0.52 on weekdays and 0.49 on weekends). The
weekend estimate stays below its own frontier of 0.145 only because 48 windows
are too few: they detect a PDI with 80\% power at the 1\% level only above
$\theta_{80}=0.21$ (Corollary~\ref{cor:mde}), and at the weekday value the test
has 46\% power. Weekend windows therefore carry the same conditional
increment as weekdays, measured with less precision.

\begin{table}[t]
\centering
\TableFont
\caption{Clock audit of the premarket-conditioned test. $S$ contains an
intercept, the Bitcoin return, and mark age; $E$ is the premarket proxy.
$\PDI_{E\mid S,T}$ is the reverse increment of the proxy given the token, and
$\theta_{80}$ the smallest PDI detectable at the 1\% level with 80\% power
(Corollary~\ref{cor:mde}). The last two rows split the sample by calendar; a
token-by-weekend interaction has HAC $t=-0.14$.}
\label{tab:timing}
% Generated by scripts/build_arxiv_assets.py from archived results; do not edit.
\setlength{\tabcolsep}{3.5pt}
\begin{tabular*}{\textwidth}{@{\extracolsep{\fill}}lrrrrrcr@{}}
\toprule
Clock or sample variant & $N$ & $\mathrm{PDI}_{T\mid S,E}$ & $c_{.01}$ & $\theta_{80}$ & HAC $t$ & Clears & $\mathrm{PDI}_{E\mid S,T}$\\
\midrule
Baseline (mark by 09:25) & 218 & 0.157 & 0.031 & 0.052 & 4.61 & yes & 0.082\\
\addlinespace
Token mark by 09:00 & 214 & 0.140 & 0.031 & 0.053 & 4.36 & yes & 0.092\\
BTC end mark by 09:25 & 106 & 0.168 & 0.064 & 0.102 & 4.46 & yes & 0.025\\
Both marks by targets & 215 & 0.170 & 0.031 & 0.052 & 5.25 & yes & 0.072\\
Marks within 30 min & 217 & 0.168 & 0.031 & 0.052 & 5.18 & yes & 0.074\\
Winsorized at 1\%/99\% & 218 & 0.165 & 0.031 & 0.052 & 5.13 & yes & 0.085\\
Proxy: 09:00-bar open & 217 & 0.150 & 0.031 & 0.052 & 4.45 & yes & 0.082\\
\addlinespace
Weekdays only & 170 & 0.122 & 0.040 & 0.066 & 4.49 & yes & 0.081\\
Weekend or holiday & 48 & 0.112 & 0.145 & 0.208 & 2.24 & no & 0.203\\
\bottomrule
\end{tabular*}

\end{table}

\subsection{Sensitivity and Closure Length}

Table~\ref{tab:sensitivity} changes the clock and the sample within the broader
Bitcoin-and-age specification. Lag limits of 30 to 60 minutes, the exclusion of
early-close starts, and winsorization leave the PDI between 0.613 and 0.629.
Requiring all four endpoints to precede their targets cuts the sample to 67
windows, because Bitcoin marks often sit seconds after their targets, and still
gives 0.445 against a frontier of 0.101. Splitting at the January 26, 2026
vintage of CoinGecko's aggregation methodology gives 0.634 before and 0.559
after. Without the premarket proxy, the increment is largest across closures:
0.784 for multi-day closures and 0.824 for the 41 Friday-to-Monday windows,
when Bitcoin and mark age say least about the listed stock's weekend move.
Section~\ref{sec:res-clock} shows that this weekend advantage comes from the
controls rather than from the token: once Monday's premarket price is included,
the token's increment is the same on weekends and weekdays.

\begin{table}[t]
\centering
\TableFont
\caption{Clock, sample, and closure sensitivity of the Bitcoin-and-age
specification. The 480-minute lag row admits one window with a stale start mark
and is reported for completeness.}
\label{tab:sensitivity}
% Generated by scripts/build_arxiv_assets.py from archived results; do not edit.
\setlength{\tabcolsep}{4pt}
\begin{tabular*}{\textwidth}{@{\extracolsep{\fill}}lrrrrr@{}}
\toprule
Check & $N$ & $R^2_0$ & PDI & $c_{.01}$ & HAC $t$\\
\midrule
\multicolumn{6}{@{}l}{\textit{Clock and filter}}\\
\quad Both token marks by targets & 219 & 0.786 & 0.629 & 0.030 & 12.12\\
\quad All four endpoints by targets & 67 & 0.781 & 0.445 & 0.101 & 5.93\\
\quad Endpoint lags at most 30 min & 221 & 0.785 & 0.629 & 0.030 & 11.71\\
\quad Endpoint lags at most 480 min & 223 & 0.785 & 0.528 & 0.030 & 5.80\\
\quad Excluding early-close starts & 219 & 0.785 & 0.621 & 0.030 & 10.81\\
\quad Winsorized at 1\%/99\% & 222 & 0.785 & 0.613 & 0.030 & 10.72\\
\addlinespace
\multicolumn{6}{@{}l}{\textit{Reference-price vintage}}\\
\quad Before Jan.\ 26, 2026 & 132 & 0.715 & 0.634 & 0.051 & 7.46\\
\quad Jan.\ 26, 2026 onward & 90 & 0.856 & 0.559 & 0.075 & 9.80\\
\addlinespace
\multicolumn{6}{@{}l}{\textit{Closure length}}\\
\quad One-day closures & 173 & 0.801 & 0.538 & 0.039 & 8.64\\
\quad Multi-day closures & 49 & 0.778 & 0.784 & 0.138 & 10.00\\
\quad Spanning a weekend & 47 & 0.777 & 0.787 & 0.145 & 10.03\\
\quad Friday to Monday & 41 & 0.769 & 0.824 & 0.166 & 9.61\\
\bottomrule
\end{tabular*}

\end{table}

Resampling checks address concentration rather than generality. In the primary
sample, leave-one-out PDIs range from 0.590 to 0.635, and none of the 221
nontrivial circular shifts of the token series produces a PDI above 0.043. For
the 41 Friday-to-Monday windows, circular-block bootstrap 95\% intervals run
from 0.664 to 0.919 with three-window blocks and from 0.683 to 0.918 with
six-window blocks, and leave-one-out PDIs range from 0.793 to 0.859. The
circular-shift $p$-value of 0.024 equals $1/41$, the smallest value this test can
attain, so a 1\% claim is not available from the subsample alone.

\subsection{Venue Construction and Execution}

Gate's trade-level mark reproduces the result on a named venue. With Bitcoin and
age only it has $\PDI=0.792$; ending the Gate return at 08:59 ET to align it
with the proxy, it clears the premarket-conditioned frontier with $\PDI=0.252$
(HAC $t=6.85$; Table~\ref{tab:main}). The Gate and composite returns have
correlation 0.976 on identical windows, with a median absolute difference of
0.30\%. At common timestamps, each source adds only a small increment to the
other ($\PDI=0.100$ for Gate given the composite and 0.057 for the reverse;
Table~\ref{tab:validation}), so the two sources largely carry the same
information. A 09:25 Gate trade gives $\PDI=0.518$ against
the proxy, but that comparison also credits about 24 minutes of extra freshness
and is not a same-clock test.

\begin{table}[t]
\centering
\TableFont
\caption{Venue construction, execution cost, and temporal validation. Order-book
costs are medians across 24 endpoint snapshots for simulated market orders.
Expanding-window and July evaluations use the Bitcoin-and-age specification;
MSE changes are relative to the same specification without the token.}
\label{tab:validation}
% Generated by scripts/build_arxiv_assets.py from archived results; do not edit.
\begin{tabular*}{\textwidth}{@{}>{\raggedright\arraybackslash}p{.36\textwidth}@{\hspace{.02\textwidth}}>{\raggedright\arraybackslash}p{.33\textwidth}@{\hspace{.02\textwidth}}>{\raggedright\arraybackslash}p{.27\textwidth}@{}}
\toprule
Check & Finding & Design and comparison\\
\midrule
\multicolumn{3}{@{}l}{\textit{Venue construction and execution}}\\
Same clock: Gate vs.\ composite at composite times & PDI 0.100 / 0.057 (Gate / composite); corr.\ 0.984 & $N=218$; $S$ adds the competing mark\\
Gate at 08:59; $S$: BTC, age & PDI 0.681 & $N=221$\\
Gate at 09:25; $S$: BTC, age, premarket & PDI 0.518 & $N=220$; trade 23.7 min after the proxy bar\\
Gate order book (\$1k) & buy/sell cost 23.7/20.8 bp & 24 snapshots; median spread 19.1 bp\\
Gate order book (\$10k) & buy/sell cost 72.0/49.3 bp & min.\ two-sided depth within 100 bp: \$436\\
\addlinespace
\multicolumn{3}{@{}l}{\textit{Temporal validation of the BTC-and-age specification}}\\
Expanding window, start $N=60/80/100$ & MSE $-59.8$/$-57.6$/$-61.3$\% & Clark--West $t=5.46$/4.79/4.43\\
July 2026, model frozen on June 5 & signs 21/22; MSE $-77.8$\% & raw token and BTC-age model: 20/22, 20/22\\
July 2026, RMSE (log points) & frozen model 0.0084; raw token 0.0081 & BTC-age model 0.0178\\
July 2026, frozen Gate model & agrees with composite 21/21; signs 20/21 & MSE $-88.1$\%; HAC $t=1.54$\\
\bottomrule
\end{tabular*}

\end{table}

Statistical informativeness is not executable value. Across the 24 endpoint
snapshots, the median quoted spread is 19.1 basis points and simulated market
orders cost 23.7/20.8 basis points to buy/sell \$1,000 and 72.0/49.3 basis points
for \$10,000; in the thinnest snapshot one side of the book held only about
\$436 within 100 basis points of the midpoint. The mark is therefore an informative measurement input
for monitoring, but not evidence that size could be traded at the mark off
hours.

\subsection{Temporal Validation}

Expanding-window forecasts that start with 60, 80, or 100 windows reduce
out-of-sample mean squared error by 59.8\%, 57.6\%, and 61.3\% relative to the
Bitcoin-and-age model (Clark--West $t=5.46$, 4.79, and 4.43
\cite{clark2007approximately}). The token's information also persists after
the estimation window. In July 2026, coefficients frozen on June 5 match 21 of
22 realized gap signs and reduce mean squared error by 77.8\% relative to the
frozen Bitcoin-and-age model, and the raw MSTRx return alone cuts that model's
root mean squared error by a similar margin (0.0081 against 0.0178; the frozen
model attains 0.0084). The holdout thus confirms the token's out-of-sample
value without separating the fitted model from the raw token, as expected when
the token loading is close to one (0.91). A frozen Gate model agrees with the
composite direction in all 21 windows covered by Gate's official July archive
and matches 20 realized signs. These evaluations concern the Bitcoin-and-age
specification; a premarket-conditioned holdout requires a later premarket
sample.

\subsection{Portability Across Products}

\begin{figure}[t]
\centering
\includegraphics[width=\textwidth]{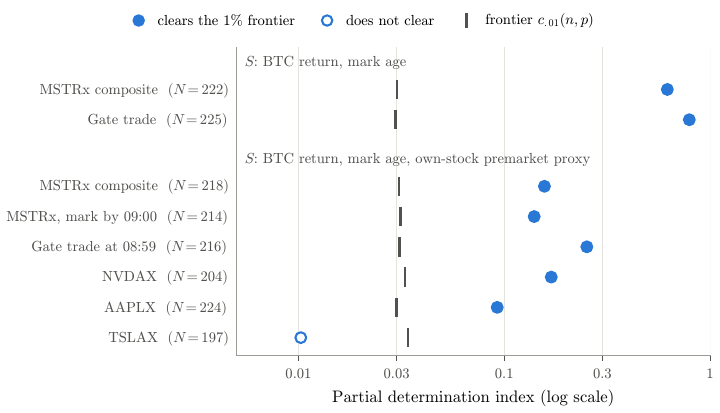}
\caption{Frontier decisions on a common scale. Each dot is an estimated PDI and
each bar the 1\% frontier $c_{.01}(n,p)$ of Theorem~\ref{thm:frontier} for
that sample; filled dots clear it. The upper group conditions on Bitcoin and
mark age, the lower group also on the own-stock premarket proxy.}
\label{fig:frontier}
\end{figure}

Figure~\ref{fig:frontier} places the frontier decisions on one scale. NVDAX ($\PDI=0.169$) and AAPLX (0.092) clear their frontiers, whereas
TSLAX does not ($\PDI=0.010$ against 0.034; HAC $t=0.80$). For these non-crypto
names, Bitcoin and age explain little of the gap, so the token's Bitcoin-and-age
PDIs are large (0.836, 0.759, and 0.717), but the own-stock premarket proxy
absorbs most of that fit. The protocol thus travels across products while
allowing its answer to differ: admission is product specific.

\subsection{From an Admitted Mark to Stack Exposure}
\label{sec:res-stack}

Table~\ref{tab:stack} turns the admitted mark into the exposure diagnostics of
Section~\ref{sec:stack}. Book leverage is 1.19 on reported first-quarter 2026
debt and 1.15 after the announced May 25, 2026 note retirement; the premium
beta of 0.103 is imprecise (standard error 0.084). The closed-window slope of
MSTRx on Bitcoin in the strict sample is 1.077, close to book leverage and well
below the 60-day listed-session beta of 1.474. For a two-times LETF these slopes
imply one-reset Bitcoin wipeout moves between $-46.4\%$ and $-33.9\%$. The
dispersion is the point: the monitored risk depends on which link of the
BTC--MSTR--LETF chain the admitted mark is used to update, and the threshold is
an exposure diagnostic, not a return forecast.

\begin{table}[t]
\centering
\TableFont
\caption{Stack calibration. Each slope estimate is mapped into the one-reset
Bitcoin move that sets a two-times LETF to zero, $-1/(2\kappa)$
(Corollary~\ref{cor:wipe}). Balance-sheet inputs use reported Bitcoin holdings
and total debt; listed-session estimates use daily data through June 5, 2026.}
\label{tab:stack}
% Generated by scripts/build_arxiv_assets.py from archived results; do not edit.
\setlength{\tabcolsep}{4pt}
\begin{tabular*}{\textwidth}{@{\extracolsep{\fill}}lrrrl@{}}
\toprule
Slope estimate & $\kappa$ & s.e. & Wipeout & Link\\
\midrule
Book leverage, Q1 2026 debt & 1.191 & -- & $-42.0\%$ & balance sheet\\
Book leverage, post-retirement debt & 1.150 & -- & $-43.5\%$ & balance sheet\\
Book leverage plus premium beta & 1.294 & 0.084 & $-38.7\%$ & plus premium\\
MSTRx closed-window slope ($N=222$) & 1.077 & 0.043 & $-46.4\%$ & admitted mark\\
60-day listed-session beta & 1.474 & -- & $-33.9\%$ & listed session\\
\bottomrule
\end{tabular*}

\end{table}

%======================================================================
\section{Discussion}
\label{sec:discussion}
%======================================================================

The central finding is not the high raw association between MSTRx and the next
MSTR open. Once shared timing and the listed stock's own premarket price are
taken into account, most of that association is absorbed by public information,
and what remains is smaller but stable: 14.0\% to 17.0\% of the residual gap
variation across the clock variants in the upper panel of
Table~\ref{tab:timing}. This is
consistent with daily-frequency evidence that listed equities lead tokenized
trackers \cite{lin2026one}. It also sharpens the weekend evidence that token
moves anticipate openings \cite{cong2025tokenized}. The token's large weekend
advantage in the Bitcoin-and-age specification reflects how little Bitcoin and
mark age reveal about a weekend move, and Monday's premarket price absorbs that
advantage. What remains is a conditional increment of the same size on weekends
and weekdays (0.112 and 0.122; interaction $p=0.89$): the token is a steady
complement to the listed stock's own premarket price, not a weekend-only
signal.

For risk monitoring, the practical implication is a sequence rather than a
verdict. A desk that wants to use a 24/7 token price before the open should
first check that the clock cannot see the target, then measure the token against
the strongest public benchmark available at the cutoff, and only then translate
an admitted mark into exposure. The same protocol applies to other asynchronous
tokenized measurements---equity perpetual marks, tokenized funds, or
oracle-fed collateral prices---whenever a mark and its target share a return
prefix.

%======================================================================
\section{Limitations}
\label{sec:limitations}
%======================================================================

The evidence is bounded in five ways. First, the premarket proxy is coarse: an
undocumented hourly chart with zero reported volume. A timestamp-matched
consolidated quote series is the natural confirmatory benchmark, and by
Lemma~\ref{lem:noisy} a sharper proxy would lower the PDI if the current proxy
is noisy.
Second, the composite's historical constituents are not disclosed; Gate supplies
one named venue, not the whole token market. Third, the main asset is a single
crypto-treasury stock over about eleven months, and calendar subsamples are
small: the 48 weekend-or-holiday windows detect only PDIs above 0.21 at the 1\%
level with 80\% power. Fourth, the out-of-sample evaluations cover the
Bitcoin-and-age specification only; a premarket-conditioned holdout awaits a
later premarket sample. Fifth, the executable
band relies on primary-market access that most holders lack, so admission
describes a monitoring input, not an arbitrage. NEST does not identify where
information originates and does not forecast post-mark shocks.

%======================================================================
\section{Conclusion}
\label{sec:conclusion}
%======================================================================

Off-hours tokenized-stock marks can look highly informative for the next
opening gap simply because both returns share the move up to the mark. NEST
separates that shared history from conditional measurement value with a
shared-history identity, a variance-clock bound, the partial determination
index, and a finite-sample frontier, and it attaches product, provenance,
timestamp, and execution audits to the statistical decision. Applied to MSTRx,
it finds that public information absorbs most of the raw fit while a modest
premarket-conditioned increment remains on the composite, on Gate, and for two
of three other xStocks. MSTRx is therefore a complementary off-hours measurement
input for the BTC--MSTR--LETF chain, and NEST is a reusable protocol for deciding
when an asynchronous tokenized mark may enter a risk system.

\declaration{Relation to the conference version.}A shorter version of this
paper has been accepted for publication in the MARBLE 2026 proceedings (Lecture
Notes in Operations Research, Springer Nature). This extended version adds the
information-clock, conditional-fit, and frontier figures; a clock audit of the
premarket comparison, including the tests without token marks later than 09:00
and with Bitcoin marks at or before the cutoff, and a calendar split with an
interaction test (Table~\ref{tab:timing}); the power function of the PDI test
and the minimum detectable PDI (Theorem~\ref{thm:frontier},
Corollary~\ref{cor:mde}); the full sample-construction, sensitivity, and validation tables; the stack
calibration and wipeout thresholds based on the strict-sample token slope
(Table~\ref{tab:stack}); proofs of all results; and an expanded literature
review. It also corrects the description of the Bitcoin data source and one
table entry of the conference version.

\declaration{Data and code availability.}Derived results and the analysis
code are available from the authors on reasonable request. Vendor
data must be obtained from the providers and cannot be redistributed.

\declaration{Disclaimer.}The views expressed are those of the authors and not
of their affiliated institutions. This paper is not investment advice.

\declaration{Competing interests.}The authors have no competing interests to
declare.

\declaration{Use of AI-assisted tools.}An AI agent assisted with drafting and
editing the extended text, generating tables and figures from archived results,
and consistency checks. The authors reviewed all output and take full
responsibility for the manuscript.

\bibliographystyle{splncs04}
\bibliography{references}

\end{document}